\documentclass[aps,prl]{revtex4}
\usepackage{amsmath}
\usepackage{amsfonts}

\begin{document}

\title{Gauge transformations are canonical transformations, redux}

\author{Z.K. Silagadze}
\email{silagadze@inp.nsk.su}
\affiliation{Budker Institute of Nuclear Physics, 630 090, Novosibirsk,
Russia and Department of physics, Novosibirsk State University, 630 090,
Novosibirsk, Russia}

\begin{abstract}
In this short note we return to the old paper by Tai L. Chow (Eur.\ J.\ 
Phys.\  {\bf 18}, 467--468 (1997)) and correct its erroneous 
final part. We also note that the main result of that paper, that gauge 
transformations of mechanics are canonical transformations, was known 
much earlier.
\end{abstract}

\maketitle 

\section{Introduction}
The letter of Tai L. Chow \cite{1} generated quite a wake in the literature
\cite{2,3,4,5}, partly because its final part is erroneous, unfortunately.
These errors, as well as other parts of the paper \cite{1}, are reproduced 
in Portuguese in \cite{2} (by the other author), and they became a subject of
consideration in two eprints \cite{3} and \cite{4}. Nevertheless, we feel
some comments are still necessary as the error was not fixed. It is the aim
of this short note to correct the proof of \cite{1} that gauge 
transformations of mechanics are canonical transformations.  

\section{Gauge transformations as canonical transformations}
It is well known that for a dynamical system two Lagrangians are equivalent
if they differ by a total time-derivative of any function of generalized 
coordinates and time:
\begin{equation}
L^\prime(q,\dot{q},t)=L(q,\dot{q},t)+\frac{d}{dt}f(q,t).
\label{eq1}
\end{equation}
The relation (\ref{eq1}) generates the following transformation of generalized
coordinates and momenta:
\begin{eqnarray}
Q_i&=&q_i,\nonumber \\
P_i&=&\frac{\partial L^\prime}{\partial \dot{q}_i}=p_i+\frac{\partial f(q,t)}
{\partial q_i},
\label{eq2}
\end{eqnarray}
because
\begin{equation}
\frac{d}{dt}f(q,t)=\sum\limits_k\dot{q}_k \frac{\partial f(q,t)}{\partial q_k}+
\frac{\partial f(q,t)}{\partial t}.
\label{eq3}
\end{equation}
It is claimed in \cite{1} that the transformation (\ref{eq2}) is a canonical
transformation, that is it preserves the Hamilton's equations:
\begin{eqnarray}
\dot{Q}_i=\frac{\partial H^\prime(Q,P,t)}{\partial P_i},\nonumber \\
\dot{P}_i=-\frac{\partial H^\prime(Q,P,t)}{\partial Q_i},
\label{eq4}
\end{eqnarray}
where the new Hamiltonian $H^\prime(Q,P,t)$ is related to the old one, 
$H(q,p,t)$, in the following way
\begin{equation}
H^\prime(Q,P,t)=\sum\limits_i P_i\dot{Q}_i-L^\prime=
H(Q,p,t)-\frac{\partial f(Q,t)}{\partial t},
\label{eq5}
\end{equation}
and here, according to (\ref{eq2}),
\begin{equation}
p_i=P_i-\frac{\partial f(Q,t)}{\partial Q_i}.
\label{eq6}
\end{equation}

It should be noted that \cite{1} was not the first paper where it was stated
that the transformation (\ref{eq2}), induced by (\ref{eq1}), is a canonical
transformation. This fact was known long ago \cite{6} and several proofs
of it can be envisaged. 

For example, one can explicitly construct the generating function \cite{6}
\begin{equation}
\Phi(q,P,t)=\sum\limits_i q_iP_i-f(q,t),
\label{eq7}
\end{equation}
so that 
\begin{equation}
p_i=\frac{\partial \Phi(q,P,t)}{\partial q_i},\;\;\;
Q_i=\frac{\partial \Phi(q,P,t)}{\partial P_i},\;\;\;
H^\prime=H+\frac{\partial \Phi(q,P,t)}{\partial t}.
\label{eq8}
\end{equation}
Just this function, and not the function $f$, as erroneously claimed in 
\cite{1}, is the generating function of the canonical transformation
(\ref{eq2}). What this transformation is the canonical transformation
follows then from
\begin{equation}
\mathrm{det}\left (\frac{\partial^2 \Phi}{\partial q_i \partial P_j}\right )
\ne 0.
\label{eq9}
\end{equation}

Another standard way to prove that (\ref{eq2}) is a canonical transformation
is to calculate the fundamental Poisson brackets \cite{3}:
\begin{eqnarray}
\{Q_i,Q_j\}&=&\sum\limits_k \left (\frac{\partial Q_i}{\partial p_k}
\frac{\partial Q_j}{\partial q_k}-\frac{\partial Q_i}{\partial q_k}
\frac{\partial Q_j}{\partial p_k}\right )=0,\nonumber \\
\{P_i,P_j\}&=&\sum\limits_k \left (\frac{\partial P_i}{\partial p_k}
\frac{\partial P_j}{\partial q_k}-\frac{\partial P_i}{\partial q_k}
\frac{\partial P_j}{\partial p_k}\right )=0,\nonumber \\
\{P_i,Q_j\}&=&\sum\limits_k \left (\frac{\partial P_i}{\partial p_k}
\frac{\partial Q_j}{\partial q_k}-\frac{\partial P_i}{\partial q_k}
\frac{\partial Q_j}{\partial p_k}\right )=\delta_{ij}.
\label{eq10}
\end{eqnarray}

Still another way, that was chosen in \cite{1}, is to directly verify the
validity of the new Hamilton equations (\ref{eq4}). Unfortunately, by 
some mysterious reason, it escaped the attention of both of the author and 
of the referee that the form of Hamilton's equations used in the final part 
of \cite{1} was erroneous thus invalidating the otherwise correct conclusions
of \cite{1}. The error was noticed in \cite{3} and \cite{4}. Reference 
\cite{3}, as was mentioned above, gave a different proof of the main claim
of \cite{1}, based on fundamental Poisson brackets, but has not corrected 
the treatment of \cite{1}. Reference \cite{4}, on the contrary, had tried
to correct the error, but, unfortunately, introducing its own mistakes, it 
came to the wrong conclusion that the main result of \cite{1} was incorrect.

In fact, the validity of the first equation of (\ref{eq4}) is not difficult 
to prove:
\begin{equation}
\frac{\partial H^\prime(Q,P,t)}{\partial P_i}=\sum\limits_k
\frac{\partial H(Q,p,t)}{\partial p_k}\frac{\partial p_k}{\partial P_i}=
\frac{\partial H(Q,p,t)}{\partial p_i}=\dot{Q}_i.
\label{eq11}
\end{equation}

It is the second equation of (\ref{eq4}) the proof of which contains some 
subtleties causing an error in \cite{4}. The subtlety is that while 
calculating the partial derivative $\frac{\partial H^\prime(Q,P,t)}
{\partial Q_i}$ it is the new momentum $P$ and not the old one $p$ which is 
held fixed. In other words, while calculating this partial derivative, we 
should take into account that $p$, according to (\ref{eq6}), is also 
a function of $Q$. Then we have
\begin{eqnarray} &&
\frac{\partial H^\prime(Q,P,t)}{\partial Q_i}=\frac{\partial H(Q,p,t)}
{\partial Q_i}+\sum\limits_k\frac{\partial H(Q,p,t)}{\partial p_k}
\frac{\partial p_k(Q,P,t)}{\partial Q_i}- \nonumber \\ &&
\frac{\partial^2 f(Q,t)}{\partial Q_i \partial t}
=-\dot{p}_i-\sum\limits_k\dot{Q}_k\frac{\partial^2
f(Q,t)}{\partial Q_k \partial Q_i}-\frac{\partial^2 f(Q,t)}
{\partial Q_i \partial t}.
\label{eq12}
\end{eqnarray}
But
\begin{equation}
\sum\limits_k\dot{Q}_k\frac{\partial^2
f(Q,t)}{\partial Q_k \partial Q_i}+\frac{\partial^2 f(Q,t)}
{\partial Q_i \partial t}=\frac{d}{dt}\frac{\partial f(Q,t)}{\partial Q_i},
\label{eq13}
\end{equation}
and we get finally
\begin{equation}
\frac{\partial H^\prime(Q,P,t)}{\partial Q_i}=-\frac{d}{dt}\left [
p_i+\frac{\partial f(Q,t)}{\partial Q_i}\right ]=-\dot{P}_i.
\label{eq14}
\end{equation}

\section{Electromagnetic gauge transformations}
Interestingly, electromagnetic gauge transformations
\begin{equation}
\vec{A}^\prime=\vec{A}+\nabla\Lambda(\vec{r},t),\;\;\;
\phi^\prime=\phi-\frac{\partial \Lambda(\vec{r},t)}{\partial t}
\label{eq15}
\end{equation}
induce the transformation of the type (\ref{eq1}) in the Lagrangian of
a classical charged particle in an electromagnetic field \cite{7}. Indeed,
if we substitute $\vec{A}\to\vec{A}^\prime,\;\phi\to\phi^\prime$ from
(\ref{eq15}) into the Lagrangian
\begin{equation}
L=\frac{1}{2}m\dot{\vec{r}}^{\;2}-e\phi(\vec{r},t)+e\dot{\vec{r}}\cdot \vec{A},
\label{eq16}
\end{equation}
we get a new Lagrangian in the form
\begin{equation}
L^\prime=L+e\frac{\partial \Lambda(\vec{r},t)}{\partial t}+e\dot{\vec{r}}
\cdot\nabla\Lambda(\vec{r},t)=L+e\frac{d\Lambda(\vec{r},t)}{dt}.
\label{eq17}
\end{equation}
As we see, the resulting transformation of the Lagrangian is of the form
(\ref{eq1}) with $f=e\Lambda$ \cite{7}.

That electromagnetic gauge transformations induce canonical transformations 
in the charged particle Lagrangian is well known \cite{6,7,8}. We see that
\cite{1}, when corrected, gives the proof of this classic result too.

\section{Concluding remarks}
That gauge transformations are a subset of canonical transformations
is, of course, known for a long time. Nevertheless, perhaps it is worthwhile 
to correct the proof of \cite{1} because it generated some confusion in the
literature. We have provided the necessary amendments in this note. Besides
we have indicated two other methods of proof. So we hope the case is settled
now: ``I have said it thrice: What I tell you three times is true'' \cite{9}.

\begin{acknowledgments}
The work is supported by the Ministry of Education and Science of 
the Russian Federation and in part by Russian Federation President
Grant for the support of scientific schools  NSh-2479.2014.2 and by
RFBR grant 13-02-00418-a.
\end{acknowledgments}

\end{document}